\documentclass[vopt,a4paper,11pt]{article}

\usepackage[utf8]{inputenc}
\usepackage[english]{babel}
\usepackage[OT1]{fontenc}
\usepackage{amsmath}
\usepackage{amsfonts}
\usepackage{amssymb}
\usepackage{graphicx}
\usepackage{geometry}

\begin{document}

\title{The conductivity of the half filled Landau level}
\author{ S.V.Iordanski\\
Landau Institute for Theoretical Physics RAS 142432 Russia,Chernogolovka}

\maketitle

\begin{abstract}
It is shown that the thermodynamic instability at the half filling of Ll leads to the vortex
lattice formation with the electronic spectrum analogous to that of graphene with two Dirac Fermi
points on Brillouin cell boundary. This result is used for the explanation of the observed current
generated by SAW in the heterostructure on the surface of piezoelectric GaAs. Using the existence
of two Fermi points instead Fermi surface suggested in the previous theoretical works, permit the
explanation of the experimental results.
\end{abstract}

The theory of electron states with Fractional Hall Effect developed in a short time after the
experimental discovery (Tsui,Stoermer,Gossard ,1982) was used mostly the projection method assuming
that the states must be constructed exclusively from the states of the first Ll like the famous
Laughlin function \cite{1},\cite{2} for $1/3$ fraction. Indeed there is no explicit calculation of the possible fractions using this assumption. The phenomenological theory of "composite" fermions \cite{3} assuming that the electrons are "dressed" by some additional magnetic flux gives some part
of the observed electron densities \cite{3} as well as a bit more refined theory of Chern-Simons
field \cite{4}.

Later it was shown in the work \cite{5} that the electron states with the partially filled Ll are
thermodynamic unstable due to the formation of the quantized vortices lowering the electron free
energy in the external magnetic field. In spite of triviality in the argumentation of \cite{5} the
result is quite general and we repeat it. The electron free energy in the external magnetic field
\cite{11} is $F=E-\int \vec{A}\vec{j}d^2r$ where $E$ is the internal electron energy, $\vec{A}$ is
the external vector-potential, $\vec{j}$ is the electrical current density. Suppose we know the internal energy $E$ at $\vec{j}=0$ which can be obtained by the minimization of the average for the
electron hamiltonian. In any case $\frac{\delta E}{\delta\psi}=0$ where $\psi$ is the electron wave
function at zero temperature. Now we change the electron wave function. Irrespective to the details
of the hamiltonian the variation of $E$ will be of the second order in the variation of the wave
function. But the change of the average current is the quantity of the first order in this variation and we can minimize the free energy creating a nonzero current. It is quite evident for the free noninteracting electrons. This phenomenon has a close analogy in the
formation of vortices in a rotating vessel \cite{6} with liquid $He_3$. In a large enough sample the vortices form the periodic vortex lattice.It is a kind of a phase transition. We shall consider strong magnetic fields when the kinetic energy term is dominating compare to Coulomb interaction term.

 The lattice periodicity in magnetic field is not enough to have a band energy structure for
electrons because the translations change the hamiltonian and one must use so called ray representations of the  periodic space groups. The simple energy band structure arise only for the
rational magnetic flux per unit cell of the vortex lattice $l/n \Phi_0$,where 
$\Phi_0=\frac{2\pi{e}}{c\hbar}$ is the unit of the flux ,($l,n)$ are co prime numbers. The requirement defining the electron density for the filled energy bands (\cite{5}) has the form
\begin{equation}
B s+K\Phi_0=\frac{l}{n}\Phi_0
\label{bsk}
\end{equation} 

where $B$ is the external  magnetic field, $s$ is the unit cell area, $K$ is the vortex circulation
number. The preference has $|K| =1$ for each individual vortex because it gives the minimal electron energy. The electron density of the fully  filled vortex lattice is
\begin{equation}
n_e=\frac{B}{\Phi_0}\frac{l-n K}{n}
\end{equation}
That covered all observed fractions in FQHE. Unfortunately there is no attempts to observe the vortex
lattice directly. It is a difficult experimental problem.

But some specific properties of the electrons in the vortex lattices can be observed not only
by FQHE which require the energy gap at the boundary of the filled band. There are the specific cases when
the vortex velocity is fully compensating the external magnetic field e.g.
\begin{equation}
Bs-2\Phi_0=0
\end{equation}
It is easy to see that this case corresponds to the half filled Ll density. This equation corresponds
to two vortices with the unit circulation in unit cell. There are also other cases with the larger
number of the vortices in unit cell but we restrict our consideration by this case in connection
with the experimental results \cite{7}. 

On the surface of the piezoelectric $GaAs$ was constructed a
high quality heterostructure $Al_xGa_{1-x}As/GaAs$. In the volume of piezoelectric $GaAs$ was generated SAW(surface acoustic wave) which produced the electric field acting on 2DES at strong magnetic field. The finite conductivity of 2DES gives the additional dissipation which can be registered to measure the conductivity at various electron densities . The results show the ohmic conductivity
at the 1/2 and 1/4 of Ll fillings. The strongest conductivity is at 1/2 filling. The experimental
results in the full extent were not explained  by the theory of the composite fermions \cite{3}, or by Chern-Simons field \cite{4}, both suggesting the existence of the Fermi surface.
We shall try to construct the physical picture and calculate the proper conductivity in the model of the vortex lattice.
We can choose the gauge with $\vec{A}_{eff}\equiv\vec{A}{eff,y}=\vec{A}_{ext}+\vec{A}_{v}$ where the effective
vector-potential is the sum of the external vector potential and the contribution of vortices 
\cite{5}. If the total flux through the unit cell vanish the magnetic translations transform into en
ordinary abelian group of translations. The main term in the hamiltonian with a strong magnetic field
shall be
\begin{equation}
H=\int \hat\psi^{+}\frac{\hbar^2}{2m}\left[-i\vec{\nabla}-\frac{e}{c\hbar}\vec{A}_{eff,y}\right]^2\hat\psi d^2r
\end{equation}
The Furrier transformation of the periodic function $A_{eff,y}(\vec{r})$ defines the reciprocal lattice. We suppose the simplest hexagonal lattice for the vortices with the simplest Brillouin cell
in the form of the hexagon with two nonequivalent vectors $\vec{k}_0$ and $\vec{k}'_0$ on its boundary.It means that (see e.g.\cite{8}) the space group of 2d  vortex crystal in the  vicinity of
$\vec{k_0}$ and $\vec{k}'_0$ inevitably has two dimensional representations and the gap between two
subsequent bands in these points vanish. The space group representations is well known in their vicinity.
Therefore we have not a Fermi surface as supposed in \cite{3},\cite{4} but two Fermi points 
\begin{equation}
\vec{k}_0=(k_0^x,k_0^y) ; \vec{k'}_0=(k_0^x,-k_0^y)
\end{equation}
where $k_0^x=\frac{2\pi}{3a}$ and $k_0^y=\frac{2\pi}{3\sqrt{3}a}$
here $a$ is the length of the unit cell side. The electron wave functions have two dimensional
representation 
$$
\psi^{+} = (\psi^{+}_1,\psi^{+}_2)
$$
as a line and
$$
\psi = \left( \begin{array}{c}
	\psi_1 \\
	\psi_2
\end{array} \right)
$$
as a column
well known for the electrons in graphene . But the length scale in the vortex lattice is of the order
of the magnetic length $l_B=\sqrt{\frac{c\hbar}{eB}}$ large compare to the inter atomic scale in  graphene at existing magnetic  fields. That is important because the electric field generated by
SAW in \cite{7} may have the wave length comparable to the size of the unit cell of the vortex
lattice.

The two dimensional representation of the space group near the points $\vec{k}_0,\vec{k'}_0$ gives
Dirac like spectrum at the electron chemical potential equal to the energy at these points
\begin{equation}
H_0=v_f\int\psi^{+}\hat p_l\sigma_l\psi d^2r
\end{equation}
Here $\sigma_l=(\sigma_x,\sigma_y) $ are $2\times2$ Pauli matrices,$v_f$ is a constant   defining the energy spectrum
\begin{equation} 
\epsilon=\mp v_f p
\end{equation}
Where $p_l=(\pi_l-k_{0l})$ or $p'_l=(\pi_l-k'_{0l})$ and $\pi_l$ is the local quasimomentum.

In order to find the electron conductivity one must perform the second quantization of the electron wave function putting
\begin{equation}
\psi(\vec{r},t) = \sum_{\vec{p}}
	\exp[i(\vec{k}_0+\vec{p})\vec{r}]
	\left(\psi_{-}(\vec{p})a_{-}(\vec{p},t)+
		\psi_{+}(\vec{p})a_{+}(\vec{p},t)
	\right)
\end{equation}

\begin{equation}
\psi^{+}(\vec{r},t)=\sum_{\vec{p}}
	\exp[-i(\vec{k}_0+\vec{p})\vec{r}]
	\left(\psi^{+}_{-}(\vec{p})a^{+}_{-}(\vec{p},t)+
		\psi^{+}_{+}(\vec{p})a^{+}_{+}(\vec{p},t)\right)
\end{equation}
where two-component functions $\psi_{\mp}(\vec{p}), \psi^{+}_{\mp}(\vec{p})$ are normalized and
orthogonal. The quantities $a_{\mp}(\vec{p},t),a^{+}_{\mp}(\vec{p'},t)$ give standard Fermi operators
in Heisenberg representation with the undisturbed hamiltonian $H_0$. We suppose that all states with
negative energy are filled. An analog representation is in the vicinity of $\vec{k'}_{0}$ 

The current is induced by the electrical potential of the SAW in the form
\begin{equation}
\label{potential}
\phi(y,t)=\left(\exp{(i\kappa y-i\omega t)}+\exp{(-i\kappa y+i\omega t)}\right)\phi_0
\end{equation}
where $\phi_0$ is real and $\omega(\kappa)=c_0\kappa$ where $c_0$ is the velocity of SAW.

The correspondent term in the hamiltonian is
\begin{equation}
H'=\int e\phi(y,t)\hat\rho d^2r
\end{equation}
where $\hat\rho(\vec{r})=\psi^{+}(\vec{r})\psi(\vec{r})$ is the density operator.

The value of the electrical potential is supposed to be small compare to the other terms in the total
hamiltonian
\begin{equation}
H=\int d^2r\psi^{+}(\vec{r})v_f\sigma_l p_l\psi(\vec{r})+e\int d^2r\phi(y,t)\psi^{+}(\vec{r})\psi(\vec{r})
\end{equation}
Treating the potential term as a small perturbation we can use Kubo expression for the average
current \cite{9}
\begin{equation}
<j_y(\vec{r},t)=-\frac{ie^2}{\hbar}\int _{-\infty}^t dt_1\int d^2r_1<[\hat j_y(\vec{r},t)\hat\rho(\vec{r_1},t_1)]>\phi(y_1,t_1)
\end{equation}
where $\hat j_y(\vec{r},t)$ is the current operator in Heisenberg representation with the unperturbed Hamiltonian $H_0$ and $\hat{\rho}(\vec{r}_1,t_1)$ is the same for the density operator. The square brackets denote the commutator for the two inside operators. The angle brackets denote the average
over the state with all negative energies filled.

The external electrical potential of SAW generate some transitions from the occupied states in the vicinity of $\vec{k}_0$ and $\vec{k'}_0$ to unoccupied  states with the positive energies. The most
important are the transitions from the vicinity of $\vec{k'}_0$ (negative energy)to the vicinity of
$\vec{k_0}$(positive energy) and the inverse transition from the vicinity of $\vec{k}_0$ to the
vicinity of $\vec{k'}_0$, both possible at a large enough wave vector $\kappa$ of the SAW. 

 But this model suggests the infinite life time of the created electron-hole pairs and give the
divergent time integral in Kubo formula. Indeed the electron-hole pairs have a finite life time
which can be taken into account by the exponentially decreasing factor $\exp{(\frac{t_1}{\tau})}$ with
some phenomenological electron-hole life time $\tau$ connected with the annihilation process on the impurities. That correspond to the results by so called "cross" technique taking into account the
scattering processes \cite{10}.

We can write the full expression for the average current in this approximation using the definitions
(6-11).
\begin{equation}
<j_y(0,0)>=-\frac{ie^2v_f}{h}\int_{-\infty}^0\exp{\frac{t_1}{\tau}}dt_1\int d^2r_1<[\psi^{+}(0,0)\frac
{p_y}{p}\psi(0,0),\psi^{+}(\vec{r}_1,t_1)\psi(\vec{r}_1,t_1)]>\phi(y_1,t_1)
\end{equation}
Here we use the uniformity of the current in time and space and put $r=t=0$ .

It is easy to show that there are two kinds of the processes corresponding to the different terms in
the commutator and generated by different terms in the real electric potential. Therefore it is possible
to consider only one process given by the matrix element with the transition $\vec{k'}_0\to\vec{k}_0$
with the corresponding term in the electric potential with $\kappa\ge k_{y0}-k'_{y0}$. 
The terms in the commutator defining the average current are

$$ n_y\sum_{all\vec{p}}
<a^{+}_{-}(\vec{p'})\psi_{-}^{+}(\vec{p'})a_{+}(\vec{p})\psi_{+}(\vec{p}) a^{+}_{+}(\vec{p'}_1,t_1)\psi^{+}_{+}(\vec{p'}_1)\\
 a_{-}(\vec{p}_1,t_1)\psi_{-}(\vec{p}_1)\exp{i(\vec{p}_1-\vec{p'}_1)\vec{r}_1}>$$

where $n_y$ is the unit vector in $y$ direction.

Putting this expression into the formula for the average current and performing the integrations
over $\vec{r_1}$ and $t_1$ and using the absence of the  excitations above the filled states with a
negative energy  one get for one electron close to $\vec{k'}_0$
$$
<j_y>=\frac{2e^2}{\hbar}v_f\phi_0(\frac{\omega(\kappa)\tau^2}{1+\omega^2\tau^2}|\psi_{+}^{+}
(k'_{oy}+\kappa)\psi_{-}(k'_{0y})|^2 S
$$
 The product of the normalization factors is equal to $S^{-2}$, where $S$ is the sample area and
 assuming $\omega \tau<<1$
 we get for the current generated by one electron close to $\vec{k'}_0$
$$
<j_y>=\frac{2e^2}{\hbar}v_fc_0\kappa\tau^2\frac{\phi_0}{S}
$$
This expression give the contribution to the total average current by one electron with quasimomentum close to $\vec{k'}_0$ (and $\vec{k}_0$ as well). If one takes the electron with the
other momentum $\vec{p}$ and the energy $\epsilon(p)\ll\hbar\omega(\kappa)$ the contribution will be practically the same. The increasing of the electron energy far from the critical points gives
the additional oscillation diminishing their contribution to the current and give the invalidation
of used Dirac  spectrum. Therefore the total contribution to the current can be estimated as
$$<j_y>=\frac{2e^2}{\hbar}v_f c_0 \tau^2\nu\kappa\frac{\phi_0}{S}=\frac{2e^2}{\hbar\lambda^2}v_fc_0\tau^2\kappa\phi_0$$
where $\nu = \frac{S}{\lambda^2}$ is the number of electrons in the "effective" vicinity of the critical points with a small energy. Here $\lambda$ is the averaged wave length of the electrons
in the "effective" domain near the critical points. It is impossible to have an electron-hole life
time $\tau$ shorter then $1/\omega$ that gives the maximal estimate $1/\omega=\tau$. That gives the crude estimate for the current independent on $\tau$
$$j_y=\frac{(2\pi)^3e^2}{\hbar}\kappa\phi_0$$

The direct calculation of the current for the other realization of the electrical potential
\begin{equation}
\phi(\vec{r},t)=\phi_0\left[\exp(-i\omega t) \exp(-i\vec{\kappa}\vec{r})+\exp(i\omega t \exp(i\vec{\kappa}\vec{r})\right]
\end{equation}
instead (\ref{potential}) replacing $\vec{\kappa}\to-\vec{\kappa}$, $\omega\to\omega$ we obtain the same formula 
for the current but with the negative $\kappa$ (the direction of the current is opposite).

Thus the conductivity is proportional to the wave vector of the SAW. That is confirmed by the experimental
results of \cite{7} giving the explicit linear dependence on the wave vector in a wide region beginning from some finite value of $\kappa$. It must be noted that in our consideration we have
neglected by thermal effects. That is possible by using Gibbs average in Kubo formula in a more complete consideration.
Thus we have an
essential difference in the physical situation when we have two Fermi points instead one Fermi surface.
The author express his gratitude to I.Kolokolov and E.Kats for numerous discussions on the subject.

\begin{figure}
	\centering
	\includegraphics[height=5cm]{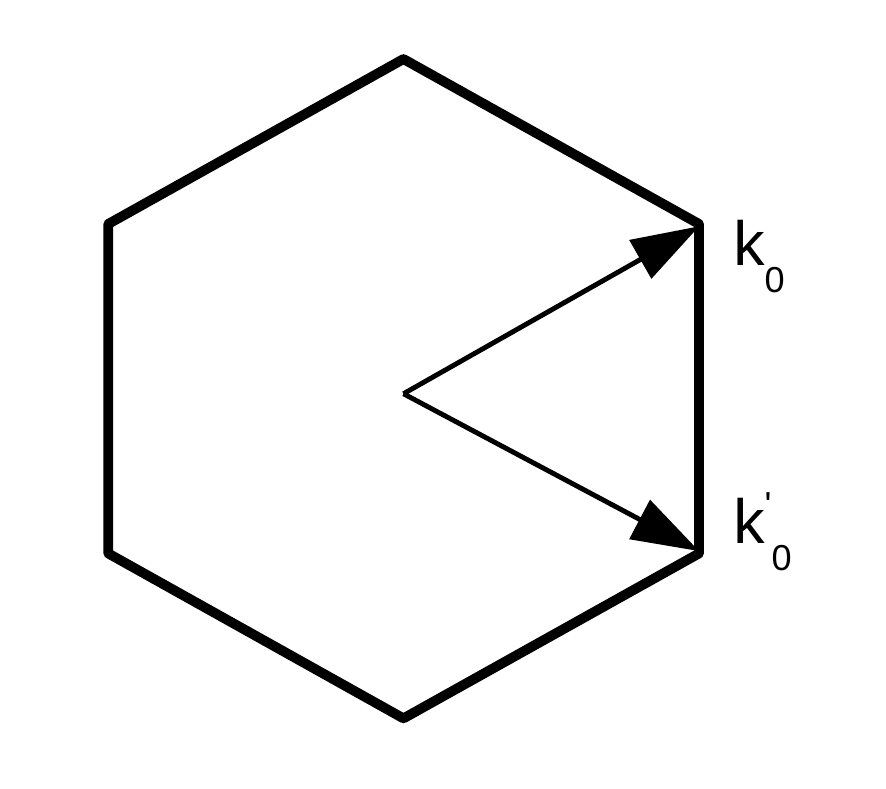}
	\caption{The Brillouin cell in the reciprocal lattice for the half-filled Ll}
\end{figure}

\begin{figure}
	\centering
	\includegraphics[height=7cm]{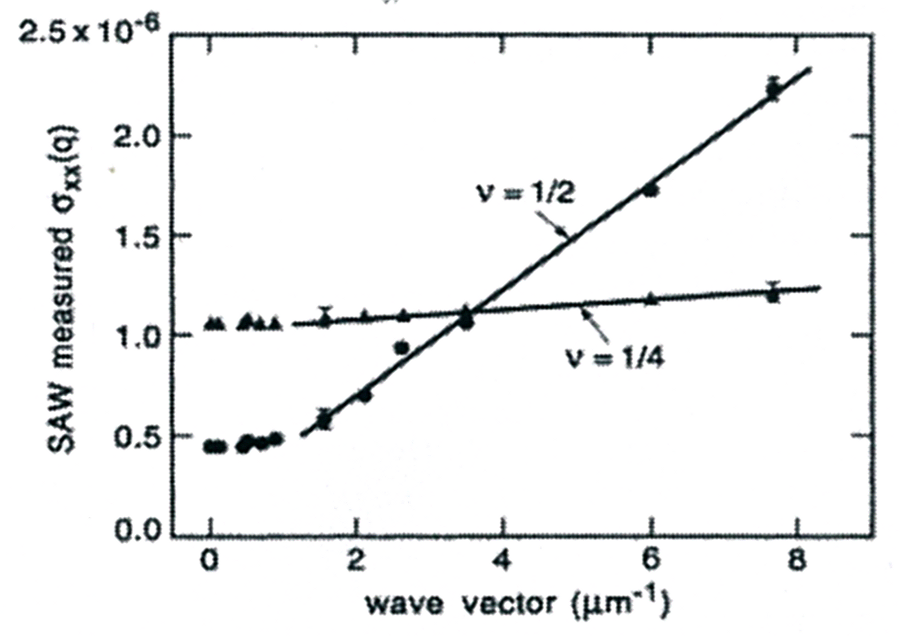}
	\caption{Reproduced from the work [7] by the permit of R.L. Willet}
\end{figure}

\end{document}